\documentclass[aps,prl,showpacs,floats,twocolumn,superscriptaddress]{revtex4}
\usepackage{amssymb}
\usepackage{amsbsy}
\usepackage{amsmath}
\usepackage{epsfig}
\usepackage{graphicx}
\usepackage{subfigure}
\usepackage{epstopdf}

\newcommand{\ibar}{\overline{\imath}}
\newcommand{\jbar}{\overline{\jmath}}

\begin{document}
\DeclareGraphicsExtensions{.eps}

\date{\today}

\title {Theory of N\'eel and Valence-Bond-Solid Phases on the Kagome Lattice of Zn-paratacamite}

\author{Michael J. Lawler}
\affiliation{Department of Physics, University of Toronto,
Toronto, Ontario M5S 1A7, Canada}

\author{Lars Fritz}
\affiliation{Department of Physics, Harvard University, Cambridge, Massachusetts 02138}

\author{Yong Baek Kim}
\affiliation{Department of Physics, University of Toronto,
Toronto, Ontario M5S 1A7, Canada}

\author{Subir Sachdev}
\affiliation{Department of Physics, Harvard University, Cambridge, Massachusetts 02138}

\date{\today}

\begin{abstract}
Recently, neutron scattering data on powder samples of Zn-paratacamite,
Zn$_x$Cu$_{4-x}$(OH)$_6$Cl$_{2}$, with small Zn concentration has been interpreted as 
evidence for valence bond solid and N\'eel ordering \cite{shlee}. 
We study the classical and quantum Heisenberg models on the distorted kagome 
lattice appropriate for Zn-paratacamite at low Zn doping. Our theory naturally leads 
to the emergence of the valence bond solid and collinear magnetic order at zero temperature. 
Implications of our results to the existing experiments are discussed. 
We also suggest future inelastic neutron and X-ray scattering experiments that can test our predictions. 
\end{abstract}


\maketitle

{\it Introduction.}--
The hallmark of frustrated magnets is the existence of 
macroscopically degenerate classical
ground states. It is believed that quantum fluctuations about such 
a highly degenerate manifold may lead to unexpected quantum 
ground states. Proposals for emergent quantum phases
include various quantum spin liquid and valence bond solid (VBS) phases
and there has been tremendous progress in a theoretical understanding 
of these phases during the last decade \cite{sachdev}. 
On the experimental front, ideal materials with spin-1/2 moments
(without orbital degeneracy) on frustrated lattices have just become
available, providing great opportunities for testing old and
new theoretical proposals \cite{hiroi,kanoda,takagi,ylee}.

One of the prime examples is
a series of recent experiments \cite{shlee,ylee,mendels,ofer,imai,harrison}
on Zn-paratacamite, 
Zn$_x$Cu$_{4-x}$(OH)$_6$Cl$_{2}$, where Cu$^{2+}$ 
ions carry spin-1/2 moments.
Most of the attention has focused on the $x=1$ limit \cite{ylee}, but
the present paper will address the $x=0$ limit \cite{shlee} when no Zn is present.
An important advantage of the $x=0$ limit is the absence of stoichiometric
disorder, which is a serious complication in the interpretation
of experiments in the $x=1$ limit.
 
At $x=1$ (herbersmithite), the idealized structure without stoichiometric disorder,
has the Cu moments residing only on the layered kagome lattices.
Remarkably, in the experiment no magnetic ordering has been found
down to 50 $mK$ even though the Curie-Weiss temperature
is $\Theta_{CW} = - 300 K$ \cite{ylee,mendels,ofer}.
This has raised the hope that the quantum ground state of
this system may be a quantum spin liquid\cite{yran,mpaf,sachdev_kagome}. 
As mentioned, however, the presence of stoichiometric disorder 
makes the interpretation of the low temperature data a difficult task. 

\begin{figure}[t]
\includegraphics[width=0.45\textwidth]{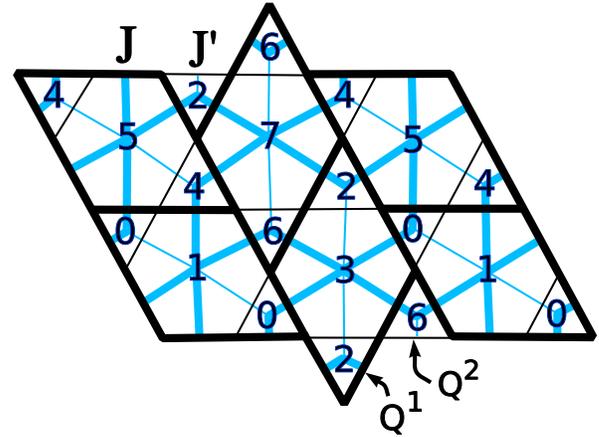}
\caption{Black (blue) represent the Distorted kagome (dual dice) lattice. 
The numbers, in multiples of 1/8, are the fractional offsets in the height model 
on the dual dice lattice which characterize the Berry's phase effects.}
\label{fig:lattice}
\end{figure}

The situation is very different near $x=0$, 
where there is no intrinsic stoichiometric disorder.
The magnetic lattice of the 
Cu$^{2+}$ spin-1/2 moments 
form stacks of alternating (distorted) 
kagome and triangular lattices.
The lattice undergoes a structural change around $x \sim 0.33$; 
monoclinic (rhombohedral) structure for $x < 0.33$ ($x > 0.33$) \cite{shlee,harrison}.
As a result, the magnetic lattice for $x < 0.33$
can be described as weakly-coupled\cite{shlee} distorted
kagome lattices (see  Fig. \ref{fig:lattice} for its structure). 

Our theory is motivated by a recent
neutron scattering experiment on powder samples of Zn-paratacamite 
for small $x$ \cite{shlee}. These experiments find two phase transitions 
at finite temperature for small $x$. In the low temperature phase, the neutron
scattering data is consistent with collinear (or N\'eel) magnetic ordering while 
no magnetic ordering is observed in the intermediate phase. Furthermore, in 
both phases a heavy gapped spin-1 mode is found as well as evidence for 
dimerization. Finally, a weak distortion in the kagome planes is observed at 
this $x$ which vanishes sharply for $x > 0.33$. Based on these results, the authors of Ref. 
\cite{shlee} propose that at small $x$ the intermediate phase is a VBS phase 
which then co-exists with magnetic ordering in the low temperature phase. It
should be noted that identification of the intermediate phase as VBS ordered 
is not consistent with the interpretation of previous $\mu$SR 
data\cite{zheng,mendels} that suggests this phase is magnetically ordered. 
As such, further experiments, preferably on single crystals, providing more 
direct measurements of VBS order in the intermediate phase is necessary to 
reconcile these experiments. It is our aim to provide useful predictions for such 
studies.

\begin{figure}[t]
\includegraphics[width=0.45\textwidth]{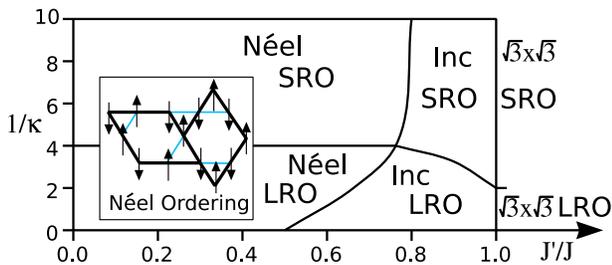}
\caption{The large-$N$ Sp($N$) mean field phase diagram. In the small 
(large) $J'/J$ regime, the N\'eel (incommensurate) long/short 
range ordered (LRO/SRO) phases arise. The inset 
represents the N\'eel ordering pattern.}
\label{fig:SpN}
\end{figure}

In the present work, we study the antiferromagnetic 
Heisenberg model on the distorted kagome lattice shown in Fig. \ref{fig:lattice}.
We consider two inequivalent 
exchange interactions, $J > J'$, consistent with the distortion
seen in the experiment, where $J$ corresponds to the shorter bond-length
(see Fig. \ref{fig:lattice}). We are mostly interested in the zero 
temperature ground states of this model and their
connection to Zn-paratacamite at $x=0$ and possibly for small $x$.
Using the Cu-O-Cu angles identified
in the experiment, one can utilize
the Goodenough-Kanamori rule
to get $J'/J \approx 0.3$ \cite{shlee2}.
In the classical Heisenberg model, we find that
energetics chooses the collinear magnetically ordered
state shown in Fig. \ref{fig:SpN} for $J'/J < 0.5$ and 
there exist highly degenerate classical ground 
states for $J' /J > 0.5$. The collinear state for $J'/J < 0.5$ 
has precisely the same magnetic order identified in the low 
temperature phase in Zn-paratacamite for small $x$ \cite{shlee}. 

\begin{figure}[t]
\subfigure[Pin-wheel state]{\includegraphics[width=0.2\textwidth]{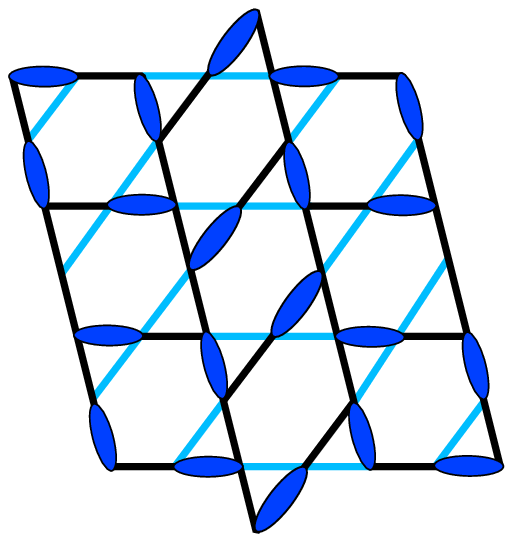}}
\subfigure[Columnar state]{\includegraphics[width=0.2\textwidth]{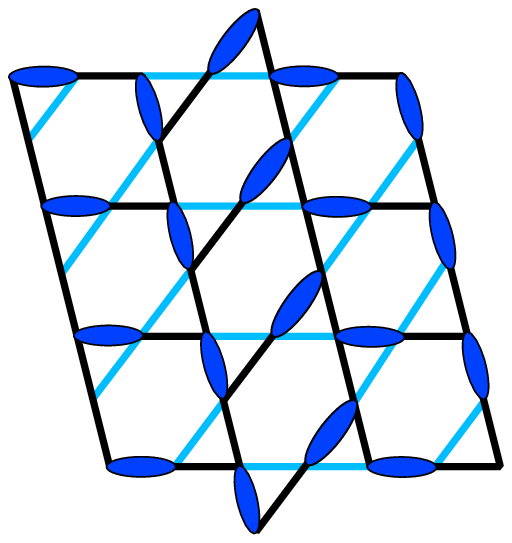}}
\subfigure[Pin-wheel triplons]{\includegraphics[width=0.22\textwidth]{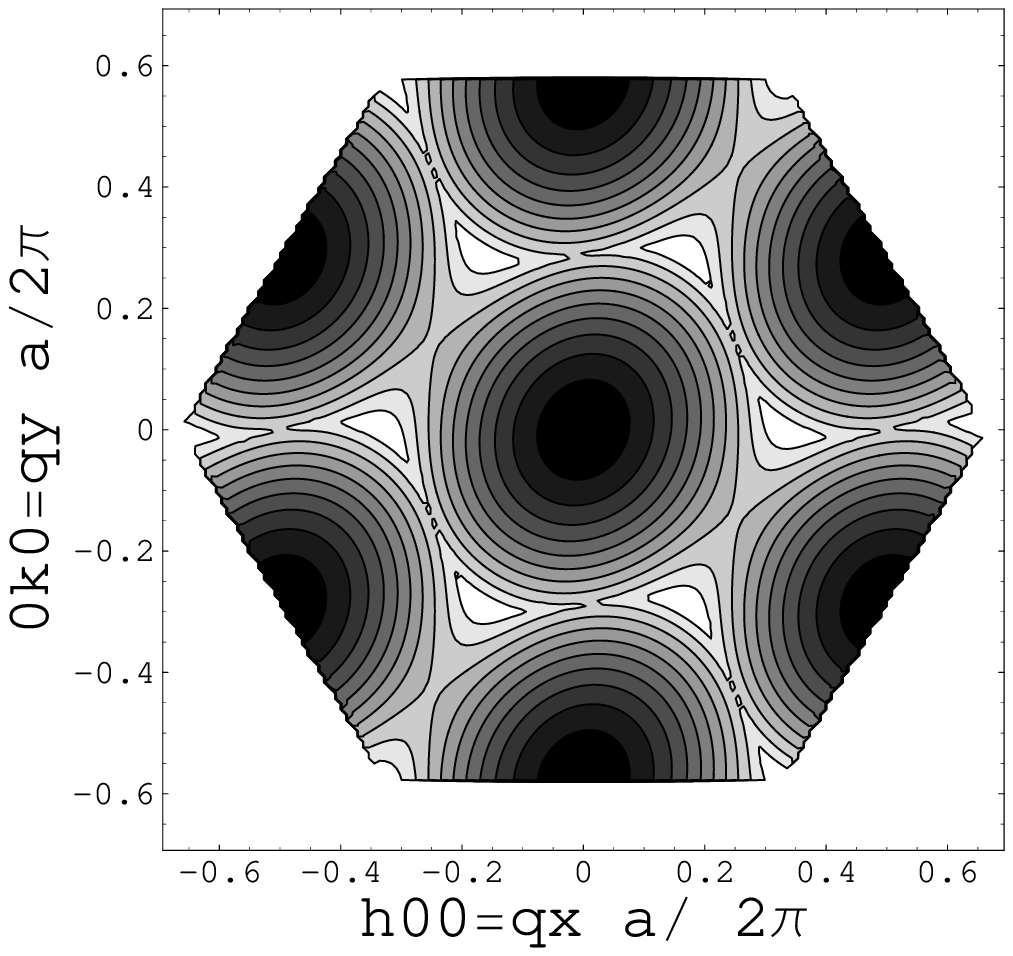}}
\subfigure[Columnar triplons]{\includegraphics[width=0.22\textwidth]{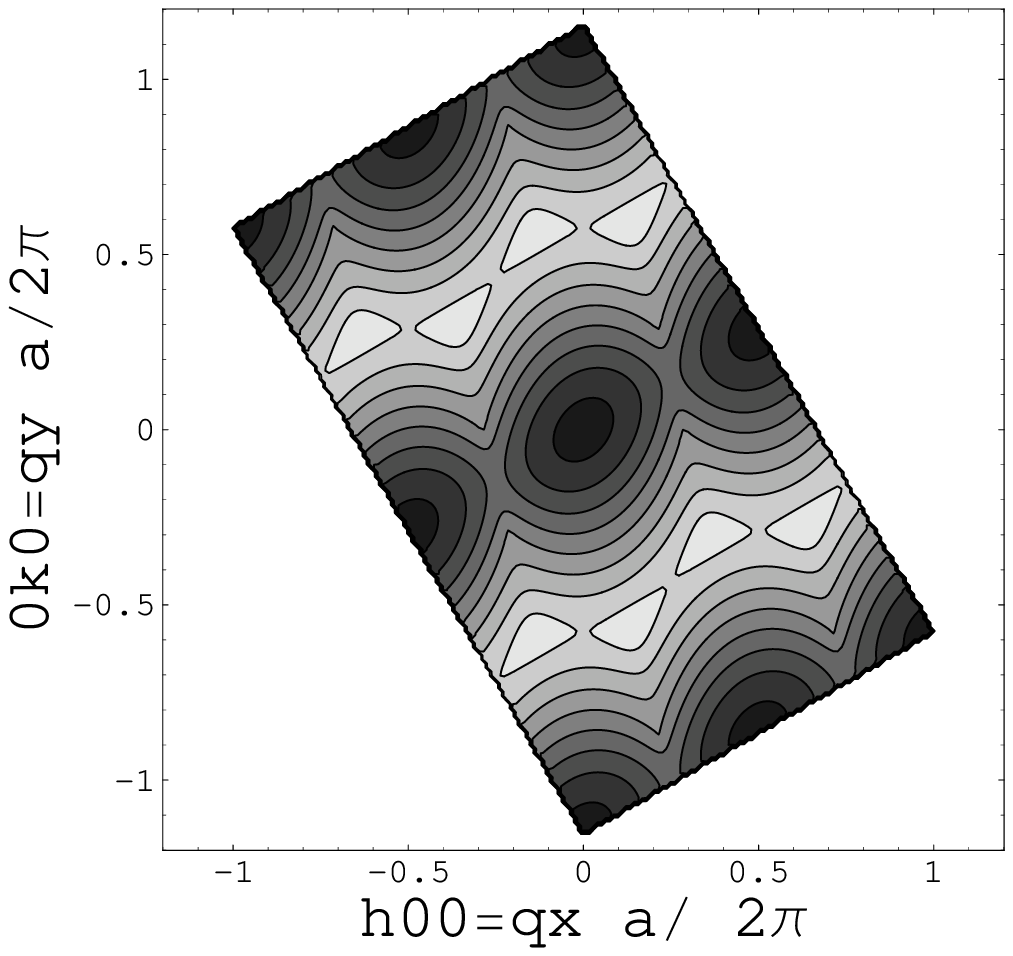}}
\caption{Pin-wheel and columnar VBS states and their corresponding lowest energy triplon excitation spectra.}
\label{fig:VBS}
\end{figure}

We then investigate
the effect of quantum fluctuations and possible 
quantum paramagnetic phases. We use the well-documented
method of the Sp(N)-generalized Heisenberg model where
one can change the magnitude of ``spin" to control
the degree of quantum fluctuations \cite{sachdev_kagome,read_sachdev}. 
The large-$N$ mean field
phase diagram of this model is presented in Fig. \ref{fig:SpN}.
Note that one gets the same collinear magnetically
ordered state for large ``spin" for $J'/J < 0.5$-$0.8$.
Understanding of the nature of the quantum paramagnetic phase for
small ``spin", however, requires careful analysis of
the spin Berry's phase and quantum fluctuation effect about
the saddle point solution \cite{read_sachdev, haldane}. We show that the resulting
quantum paramagnetic state is a VBS phase
depicted in Fig. \ref{fig:VBS}a. We call this the ``pin-wheel" VBS state. 
It is to be distinguished from the ``columnar" VBS
state (shown in Fig. \ref{fig:VBS}b) which was 
suggested as a candidate valence bond order in the 
Ref. \cite{shlee}.
It can be shown\cite{read_sachdev} that the ``pin-wheel" state can lower
the energy by the resonating moves of dimers around
the ``pin-wheel" structures.  

\begin{figure}[t]
\includegraphics[width=0.25\textwidth]{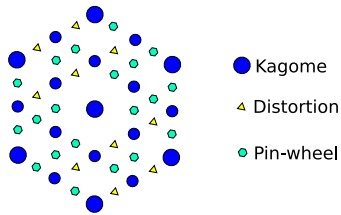}
\caption{X-ray structure factor: circles represent Bragg peaks
of the ideal kagome lattice; triangles arise from the structural 
distortion shown in Fig. \ref{fig:lattice}. These are the only Bragg 
peaks in the columnar state. In the pin-wheel state, additional
Bragg peaks (hexagons) appear due to further lattice distortion.}
\label{fig:scattering}
\end{figure}

In order to provide definite predictions for the
valence bond solid phase, we compute the triplon
dispersions (shown in Fig. \ref{fig:VBS}c-d) for both of the
VBS phases and suggest that 
inelastic neutron scattering
will be able to distinguish these phases via
their quite different triplon dispersions when
a single crystal sample becomes available. 
We also suggest that an X-ray 
scattering experiment may clearly distinguish the two
VBS ordering patterns via their different further lattice 
distortions induced by the ordering. The expected
X-ray structure factors for both phases
are shown in Fig. \ref{fig:scattering}.

{\it Classical Heisenberg Model.}--
Possible magnetic ordering patterns in the classical Heisenberg model
can be investigated by studying the O($N$) model in the large-$N$ limit \cite{canals}
where the 3 component spin unit vector is replaced by an N-component real-valued
vector $\vec\phi$, with $\vec\phi\cdot\vec\phi=N$. The collinear magnetic order
shown in Fig. \ref{fig:SpN}, the same magnetic 
order observed in the experiment, is chosen for $J'/J < 0.5$ by this method. 
%
When $J'/J > 0.5$, a highly degenerate set 
of wavevectors have the same lowest eigenvalue and 
energetics alone does not determine any particular
magnetic order. Thus, if magnetic ordering occurs for $J'/J > 0.5$ at finite
temperatures, it should arise via a thermal order by disorder
phenomenon.

{\it Quantum Sp(N) Model and Mean Field Theory.}--
To investigate possible magnetically ordered and 
quantum paramagnetic states in the quantum antiferromagnetic
Heisenberg model, it is useful to generalize the usual 
spin-SU(2) Heisenberg model to an Sp(N) model \cite{read_sachdev}.

Let us start with the Schwinger boson representation of
the spin operator
${\vec S}_r = 
b^{\dagger}_{r \alpha} {\vec \sigma}_{\alpha \beta} b_{r \beta}$,
where $\alpha,\beta = \uparrow, \downarrow$, ${\vec \sigma}$ are
Pauli matrices, $b_{r \alpha}$ are canonical boson operators and
a sum over repeated $\alpha$ indices is assumed. 
Note that we need to impose the constraint 
$n_b = b^{\dagger}_{r \alpha} b_{r \alpha} = 2S$
to satisfy the spin commutation relations, where $S$ is the spin
quantum number. A generalized model is
then obtained by considering $2N$ bosonic fields,
$b_{r \alpha}$ with $\alpha = 1, 2, ... , 2N$ and the constraint
$n_b = b^{\dagger}_{r \alpha} b_{r \alpha} = 2SN$.
The simplest such model with Sp(N) symmetry is
\begin{equation}
{\cal H} = {1 \over 2} \sum_{rr'} J_{rr'} 
({\cal J}_{\alpha \beta} b^{\dagger}_{r \alpha} b^{\dagger}_{r' \beta})
({\cal J}_{\gamma \delta} b_{r \gamma} b_{r' \delta}),
\end{equation}
where ${\cal J}_{\alpha \beta}$ is a $2N \times 2N$ totally
antisymmetric matrix which generalizes the Pauli matrix $i\sigma_2$ 
from $N=1$.
When $\kappa = n_b/N = 2S$ is fixed while taking 
the large-$N$ limit, the saddle point solution can be 
classified using both
the valence bond order parameter
$Q_{rr'}  = \langle {\cal J}_{\alpha \beta} b^{\dagger}_{r \alpha} 
b^{\dagger}_{r' \beta} \rangle$ and the magnetization induced
by a finite condensate $x_{r \alpha}  = \langle b_{r \alpha} \rangle$.
The advantage of the large-$N$ mean field theory is that
we can
investigate both the large-$\kappa$ semiclassical
limit and the small-$\kappa$ extreme quantum limit
on an equal footing\cite{read_sachdev}.

In the distorted kagome lattice, we need two inequivalent valence 
bond order parameters, $Q^1_{rr'}$ and $Q^2_{rr'}$, as depicted 
in Fig. \ref{fig:lattice}. We also need two Lagrange multipliers for two inequivalent
sites to impose the constraint $b^{\dagger}_{r \alpha} b_{r \alpha} = \kappa N$.
These parameters need to be determined self-consistently in 
the large-$N$ mean field theory.
The large-$N$ Sp(N) mean field phase diagram for the distorted
kagome lattice is shown in Fig. \ref{fig:SpN}. 
For $\kappa > \kappa_c = 0.26$, the collinear magnetically 
ordered state (Fig. \ref{fig:SpN}) appears in the $J'/J < 0.5$-$0.8$ regime; 
this is the same magnetic order as discovered in the experiment
and in the classical model. For $J'/J > 0.5$-$0.8$ and at large $\kappa$,
the ground state acquires an incommensurate coplanar order
and becomes the $\sqrt{3} \times \sqrt{3}$ state at $J'=J$ \cite{sachdev}. 
The nature of the paramagnetic state for small $\kappa < \kappa_c$, 
however, cannot fully be determined within the mean field theory.

{\it Quantum Fluctuations and Valence Bond Solid in the Paramagnetic Phase.}--
Understanding of the paramagnetic phases requires careful analysis of 
spin Berry's phase and 
quantum fluctuation effects about
the mean field solution \cite{read_sachdev, haldane}. 
It is important to note that $Q^2_{rr'}  = 0$ in the
paramagnetic phase for small $J'/J < 0.5$-$0.8$. Thus
this phase is adiabatically connected to the $J'=0$ limit, corresponding 
to the bipartite lattice depicted by the thick lines in Fig. \ref{fig:lattice}.
Using $Q^1_{rr'} = Q_1 e^{i A_{rr'}}$, one can clearly see that
the action in this $Q^2_{rr'}=0$ paramagnetic phase is invariant under 
the U(1) gauge transformation: $b_{r \alpha} \rightarrow e^{i \theta_r} b_{r \alpha}$ 
($b_{r \alpha} \rightarrow e^{-i \theta_r} b_{r \alpha}$) for 
$r$ on the A-sublattice (B-sublattice) and 
$A_{rr'} \rightarrow A_{rr'} + \theta_r - \theta_{r'}$.
The effective field theory of such a paramagnetic phase
is given by the gapped bosonic spinons carrying $\pm 1$ gauge charges 
(depending on the sub-lattices) coupled to a U(1) gauge field $A_{rr'}$.
Since the spinons are gapped in the paramagnetic phase, integrating them out 
in general produces a 2+1 dimensional  compact U(1) lattice gauge theory 
captured by the simple partition function \cite{sachdev}
\begin{equation}
Z = \int \prod_{\langle ij \rangle} {d A_{ij} \over 2\pi}
\exp [ {\sum_p V ({\rm curl}_p \  {\bf A}) + i \sum_{\langle ij \rangle} \eta_{ij} A_{ij}} ],
\end{equation}
where $\langle ij \rangle$ represent the nearest-neighbor sites of 
the space-time lattice (here we have discretized time) and
$V(\Phi) = V(-\Phi) = V(\Phi+2\pi)$ is an arbitrary periodic potential.
Here $p$ labels the plaquette of the space-time lattice and
${\rm curl}_p \ {\bf A} = \sum_{\langle ij \rangle \in p} {\rm sgn}_p(ij) A_{ij}$,
where ${\rm sgn}_p(ij) = - {\rm sgn}_p(ji) = 1$ if $j$ comes right after $i$ 
when one goes around a given plaquette $p$ and ${\rm sgn}_p(ij) = 0$ otherwise.
Here $\eta_{ij}$ is an external current determined by spin Berry's phase
and it is given by $\eta_{ij} = \eta_{(rt), (r't')} = \pm \delta_{rr'} \delta_{t+1,t'}$
(for spin-1/2) depending on whether $r$ belongs to the A- or B-sublattice.
Thus the problem reduces to the compact U(1) gauge theory with 
background charges of $\pm 1$ at the A- and B-sublattice \cite{sachdev}.
As well known, this compact U(1) gauge theory is confining and
the resulting ground state would most likely be a VBS.

In order to find the nature of the VBS state, it is
useful to construct the so-called height model on the dual lattice \cite{sachdev},
which is equivalent to the compact U(1) gauge theory on the direct lattice.
The height model can be derived using
the well-documented duality transformation and written in terms of 
the integer-valued height fields $h_{\overline{\imath}}$ defined on the 
sites ${\overline {\imath}}$ of the dual space-time lattice \cite{sachdev}.
In our case the dual lattice (in a given time slice)
is a distorted dice lattice $\{ {\bar r} \}$ as shown in Fig. \ref{fig:lattice}
(blue lattice).
Note that the thick blue lines correspond to the dual lattice 
of the $J'=0$ limit of the original distorted kagome lattice. 
The height model is found to have action
\begin{equation}
S_h = 
\sum_{\langle {\overline {\imath}} {\overline {\jmath}} \rangle} {g \over 2}
(h_{\overline{\imath}} - h_{\overline{\jmath}} + \zeta_{\overline{\imath}} 
- \zeta_{\overline{\jmath}})^2
\end{equation}
where $\langle\overline{\imath}\overline{\jmath}\rangle$ are the thick
bonds of the distorted dual space-time lattice and $g$ a non-universal 
coupling constant. Here the constraint $h_{\overline{\imath}}=h_{\overline{\jmath}}$ must
be imposed if $\overline{\imath}\overline{\jmath}$ is a thin bond and the offset 
variables $\zeta_{\overline{\imath}} $ are determined by the spin Berry's 
phase and their time independent site-dependent
values $({1 \over 8}, {6 \over 8}, {3 \over 8}, 0, {5 \over 8}, {2 \over 8}, {7 \over 8}, {4 \over 8})$ 
on the dice lattice are shown in Fig. \ref{fig:lattice}.
After solving the simple constraint, this height model can be 
understood using standard methods\cite{read_sachdev} and the 
average height fields can be determined up to an overall 
constant \cite{sachdev}. 

The nature of the VBS ground state can be
studied by using the relation between the height fields
and the VBS order parameter.
It can be shown that the ``electric field" (in the compact 
U(1) gauge theory) defined on the spatial dual-lattice links 
is related to the height fields via $e_{{\bar r} {\bar r}'}  
= \langle h_{\bar r}\rangle -\langle h_{{\bar r}'}\rangle$ \cite{sachdev}. The VBS order
parameter defined on the direct-lattice link that intersects
the spatial dual-lattice link $\langle {\bar r} {\bar r}' \rangle $ is 
given by the strength of $e_{{\bar r} {\bar r}'}$ \cite{sachdev}.
The result is a ``pin-wheel" pattern shown in Fig. \ref{fig:VBS}a.  

{\it Neutron Scattering:~Triplon Dispersion.}-- In the Ref. \cite{shlee},
a different VBS phase (see Fig. \ref{fig:VBS}b) - the ``columnar"
phase---was suggested. Here we show that different triplon
dispersions in the pin-wheel and columnar VBS
phases can be used to distinguish them if inelastic
neutron scattering experiments are done on single crystals.

To compute the triplon dispersion, consider letting $J$ be the exchange
interaction between two spins within the same valence bond and 
$\lambda J$ between two spins on different nearby valence bonds. In the
decoupled $\lambda=0$ limit, the triplon dispersion would be completely flat
with energy $J$. When $\lambda$ is finite, the triplon band  disperses.
Here we compute this dispersion to first order in $\lambda$.
For this purpose, we apply the bond-operator formalism\cite{sachdev_bond} 
to the valence bonds in the VBS phases where the Hilbert space can be
represented via singlet and triplet states on the bonds of Fig. \ref{fig:VBS}. 
At first order in $\lambda$, only the processes that preserve triplon number contribute, 
and to this order they become dispersing particles. These dispersions for the lowest 
band in both the pin-wheel and columnar states are shown in Fig. \ref{fig:VBS}c-d. 
The minima in the two cases are clearly located at 
different positions, a feature that can be distinguished experimentally.

{\it X-ray Scattering.}--
Assuming a lattice contraction where valence bonds exist, 
the pin-wheel state should break the lattice translational symmetry of the distorted
kagome lattice in one of two directions, doubling the unit cell. This would
lead to new peaks in the X-ray structure factor.
However, the lattice translational symmetry would be intact in the
columnar state, leading to no new Bragg peaks in the
X-ray structure factor in addition to those associated with the
distorted kagome lattice. The X-ray structure factors for the two VBS
phases are shown in Fig. \ref{fig:scattering}. Note that the hexagon symbols
represent the new Bragg peaks in the pin-wheel state. All other
peaks also exist in the columnar state.

{\it Summary and Conclusion.}--
We have provided a theory of the zero temperature phases of an 
antiferromagnetic Heisenberg model on a distorted kagome lattice. 
The resulting VBS and N\'eel ordered phases are strikingly similar 
to those identified in the recent neutron 
scattering experiment on Zn-paratacamite at small doping $x$ \cite{shlee}. 
In particular, our theory predicts that ``pin-wheel" VBS ordering 
(see Fig. \ref{fig:VBS}a) can occur as a result of quantum disordering 
of the N\'eel order. 
We have suggested future neutron and X-ray scattering experiments that can
test our predictions for this ``pin-wheel'' VBS ordering. 
Our predictions may also be used for future resolution of
the disagreement between the interpretations of the neutron 
scattering and $\mu$SR data in the intermediate 
temperature phase. Furthermore, here we focused on zero temperature 
ground states of a Heisenberg model
so that an explanation of the coexistence of N\'eel and VBS 
ordering at finite temperature is beyond the scope of this work. However, we 
note that a phase transition between the two phases is likely to be
first order leaving the possibility of a coexist region in the phase diagram. 
Possible relation between the quantum phases
on the distorted kagome lattice described here and the yet-to-be-determined
quantum ground state\cite{yran,mpaf,sachdev,claire,sachdev_kagome,kagometheory} on 
the ideal kagome lattice is an important subject of future research.

This work was supported by NSERC, CIFAR, CRC, and
KRF-2005-070-C00044 (MJL, YBK); NSF Grant No.~DMR-0537077
(LF, SS); Deutsche Forschungsgemeinschaft under grant FR 2627/1-1 (LF). 
We thank S.-H. Lee, A. Vishwanath and M. Hermele for helpful discussions,
and acknowledge the Aspen Center for Physics.

\renewcommand{\theequation}{S\arabic{equation}}
\renewcommand{\thefigure}{S\arabic{figure}}
\setcounter{equation}{0}  
\setcounter{figure}{0}  
\section{Supplementary Information}
\noindent{\it Electromagnetic duality on the distorted kagome lattice.}
The electromagnetic duality of a U(1) lattice gauge theory on a two dimensional square lattice is well known and documented.  Here we follow Ref. \cite{Sachdev:2002uq} for the case of the duality of a U(1) lattice gauge theory on the strong bonds of the distorted kagome lattice of undoped paratacamite. In this $2+1$ dimensional model, this duality mapping connects the U(1) gauge theory with a scalar ``height" model. An analysis of such a model is then straightforward and includes a clear picture for exactly how the confinement occurs: the spinons will bind into singlets and a valence bond solid (VBS) forms. 

Recall the U(1) lattice gauge theory of Eq. 3:
\begin{equation}
Z\! =\!\! \int \prod_{\langle ij \rangle} {d A_{ij} \over 2\pi}
\exp \bigg[ {\sum_p V ({\rm\bf curl}_p \  {\bf A}) + i \sum_{\langle ij \rangle} \eta_{ij} A_{ij}}\bigg],
\end{equation}
The first step in transforming to the dual theory is to Fourier transform the periodic function $\exp\{V(\Phi_p ={\rm\bf curl}_p \ {\bf A})\}$ on each plaquette:
\begin{equation}
  \exp V(\Phi_p) = \sum_{e_p\in{\mathbb Z}}\exp\big[\tilde V(e_p) + ie_p\Phi_p\big]
\end{equation}
where  $\exp{\tilde V(e_p)}$ is the fourier transform of $\exp{V(\Phi_p)}$ and $e_p$ are integer electric fields living on the plaquettes $p$ of the direct lattice. After performing this transformation on each plaquette, we integrate out $A_{ij}$ to obtain
\begin{equation}
 Z = \prod_p\sum_{\{e_p\in \mathbb{Z}\}}
 \delta\bigg({\rm\bf div}_{ij} {\bf e} - \eta_{ij}\bigg)
  \exp\bigg\{\sum_{p} \tilde V(e_p) \bigg\}
\end{equation}
where ${\rm\bf div}_{ij}$ is the lattice divergence of $e_p$ with plaquettes $p$ that share the bond $\langle ij\rangle$. Thus we have replaced $A_{ij}$ with the integer electric fields $e_p$. 

At this stage it is useful to switch to a description on a connected dual lattice whose bonds pierce the plaquettes of the direct lattice. Such a lattice is actually well known in the isotropic limit for the dual of the kagome lattice is the dice lattice. Fig. 1 of the main text extends both of these lattices to the distorted case.

The duality mapping is then completed by a suitable mapping that relates $e_p$ to $b_{\ibar\jbar}=-b_{\jbar\ibar}$ and $\eta_{ij}$ to $\eta_{\bar p}$, where $\ibar$ and $\jbar$ label the sites of the dual lattice of Fig. 1 that form the bond $\langle\ibar\jbar\rangle$ which pierces the plaquette $p$ of the direct lattice and $\bar p$ labels a plaquette of the dual lattice. Note: we remove the ambiguity $p\to \ibar\jbar$ or $p\to \jbar\ibar$ by associating an outward normal for each plaquette using the right hand rule on the A sublattice of the bipartite direct lattice. The resulting partition function is
\begin{equation}\label{eq:Zsij}
 Z\! = \!\!\!\sum_{\{b_{\ibar\jbar}\in \mathbb{Z}\}}\!
 \prod_{\langle\ibar\jbar\rangle\in \text{thin}}\!\!\!\!\! \delta_{b_{\ibar\jbar},0}
 \prod_{\bar p}\delta\big(
   \text{\bf curl}_{\bar p}\ {\bf b}  - \eta_{\bar p}\big)\!
  \exp\bigg[\!\sum_{\langle\ibar\jbar\rangle\in \text{thick}}\!\!\!\!\!\! \tilde V(b_{\ibar\jbar}) \bigg]
\end{equation}
Thus the Gauss' law constraint on $e_p$ in the direct description becomes Amp\`ere's law for $b_{\ibar\jbar}$ in the dual description, provided we set $b_{\ibar\jbar}=0$ on the thin bonds. Aside from this one caveat, this electro-magnetic duality transformation directly follows that on square lattice\cite{Sachdev:2002uq}.

\begin{figure}[th]
\subfigure[$b^0_{\ibar\jbar}$]{\includegraphics[width=0.22\textwidth]{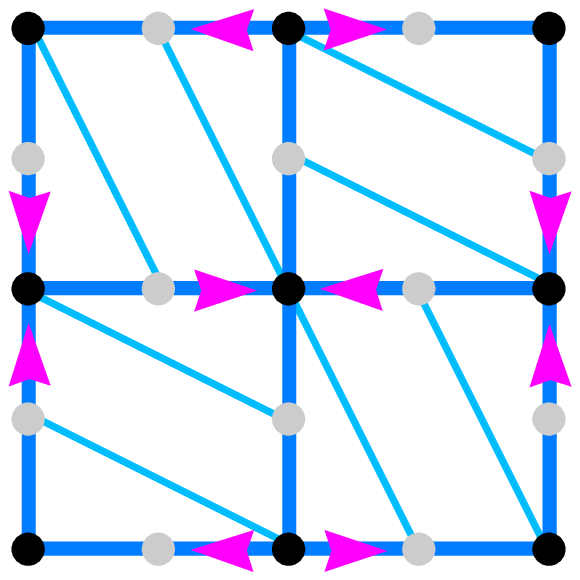}}
\subfigure[decomposition]{\includegraphics[width=0.21\textwidth]{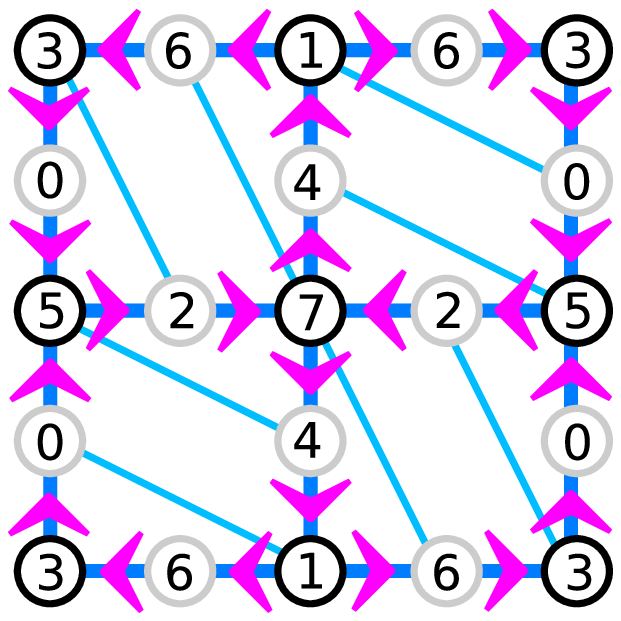}}
\caption{Solution to the Amp\'ere's law constraints on $b_{\ibar\jbar}$ on the dual lattice (reshaped for convenience). a) one choice of the particular solution with $b^0_{\ibar\jbar}=-b^0_{\jbar\ibar} = 1$ if $\ibar$ and $\jbar$ are connected by an arrow from $\ibar$ to $\jbar$; $b^0_{\ibar\jbar}=0$ otherwise. b) The decomposition into divergence and curl free parts: $H_{\ibar\jbar}=-H_{\jbar\ibar}=3/8$ if $\ibar$ and $\jbar$ are connected by an arrow from $\ibar$ to $\jbar$, $H_{\ibar\jbar}=0$ on the dashed line bonds; the fractional offsets $\zeta_{\ibar}$ are the numbers on the sites of the lattice divided by $8$.}
\label{fig:ampere}
\end{figure}

It remains to solve the constraints in Eq. \eqref{eq:Zsij} to obtain a useful representation of the dual theory.  In general, we can solve the Amp\`ere's law constraint for $b_{\ibar\jbar}$ by letting
\begin{equation}
  b_{\ibar\jbar} = b^0_{\ibar\jbar} + h_{\ibar} - h_{\jbar} 
\end{equation}
where $h_{\ibar}$ are integer ``height" fields that live on the sites of the dual lattice and the integers $b^0_{\ibar\jbar}$ satisfy
\begin{equation}\label{eq:ampere}
  {\textstyle \sum_{\langle\ibar\jbar\rangle\in \bar p} }{\bf curl}_{\bar p}\ {\bf b^{0}}  = \eta_{\bar p}
\end{equation}
which can take any one of many gauge equivalent configurations. One possible solution is shown in Fig. \ref{fig:ampere}a. Given such a solution, it is useful to split it up into its divergence free and curl free parts:
\begin{equation}
  b^0_{\ibar\jbar} = \zeta_{\ibar} - \zeta_{\jbar} + H_{\ibar\jbar}
\end{equation}
where $H_{\ibar\jbar}$ is the divergence free part and $\zeta_{\ibar}$ and $H_{\ibar\jbar}$ are fractions whose sum give the integers $b_{\ibar\jbar}$. For our solution $H_{\ibar\jbar}$ and $\zeta_{\ibar}$ are shown in Fig. \ref{fig:ampere}b. In conjunction with this decomposition it is also useful to specialize to the Villain model $\tilde V(e_p) = -\frac{g}{2}(e_p)^2$ for
\begin{multline}
\left(h_{\ibar}-h_{\jbar}\! +\! \zeta_{\ibar} - \zeta_{\jbar}\! +\! H_{\ibar\jbar}\right)^2 
    \!=\! \left(h_{\ibar}-h_{\jbar} + \zeta_{\ibar} - \zeta_{\jbar}\right)^2 -
    \frac{e^2}{2}\left( H_{\ibar\jbar}\right)^2 \\ 
    +\text{terms that vanish in the sum over $\ibar$,$\jbar$}
\end{multline}
so that $H_{\ibar\jbar}$ only contributes to an overall constant out front of the partition function. 

Thus, the solution of the Amp\`ere's law constraint leaves us with the height model of Eq. 4 of the main text defined on the sites of the dual lattice:
\begin{equation}\label{eq:Zdual}
Z = \prod_{\overline {\imath}} 
\sum_{h_{\overline {\imath}}\in \mathbb{Z}}
\left [ \prod_{\langle \overline{\imath}\overline{\jmath}\rangle \in \text{thin}} 
\delta\big(h_{\overline {\imath}} - h_{\overline {\jmath}}\big) \right ] e^{-S_h},
\end{equation}
with $S_h = 
\sum_{\langle {\overline {\imath}} {\overline {\jmath}} \rangle\in \text{thick}} {g \over 2}
(h_{\overline{\imath}} - h_{\overline{\jmath}} + \zeta_{\overline{\imath}} 
- \zeta_{\overline{\jmath}})^2$
and $g$ a non-universal coupling constant. 

{\it Confinement and the pin-wheel VBS phase.}
The remaining constraints, $h_{\ibar} - h_{\jbar} + b^0_{\ibar\jbar}=0$ if $\langle\ibar\jbar\rangle$ is a thin bond, reduce the number of independent height fields and fractional offsets $\zeta_{\ibar}$ to those living on the sites of a \emph{square lattice}. An understanding of \eqref{eq:Zdual} then proceeds by a series of mappings onto already known results.  

The dual lattice, shown in Fig. \ref{fig:ampere}, has two types of sites: black sites with four neighbors which lie on the vertices of a square lattice and gray sites with two neighbors that are decorations which lie on the bonds of a square lattice. The constraint across the thin bonds relates two decorations to a vertex. Denoting $v$ as a vertex site and $d(v)$ as one of its decorations, we have
\begin{equation}
  h_{d(v)} = h_{v}, \quad \zeta_{d(v)} = \zeta_v - \tfrac{1}{8}
\end{equation}
where $\zeta_v \in \{1/8,3/8,5/8,7/8\}$ cyclicly around a plaquette of the square lattice. Utilizing this mapping, and shifting all $\zeta_v$ by $1/8$, the height model takes the form
\begin{multline}
Z = \prod_v \sum_{h_v\in\mathbb{Z}} \exp\Bigg\{
 -\sum_{\langle vw\rangle} \frac{g}{2}\big(h_v-h_w + \zeta_v - \zeta_w\big)^2\\ - 
 2\sum_{\langle\langle vw\rangle\rangle} \frac{g}{2}\big(h_v-h_w+\zeta_v-\zeta_w\big)^2\Bigg\}
\end{multline}
with $\zeta_v = \{0,1/4,2/4,3/4\}$. This height model differs from the simple square lattice case studied by Sachdev and Park\cite{Sachdev:2002uq} by the addition of off diagonal terms that convert it to a model on the anisotropic triangular lattice.

\begin{figure}[b]
\subfigure[4-site sublattice]{\includegraphics[width=0.105\textwidth]{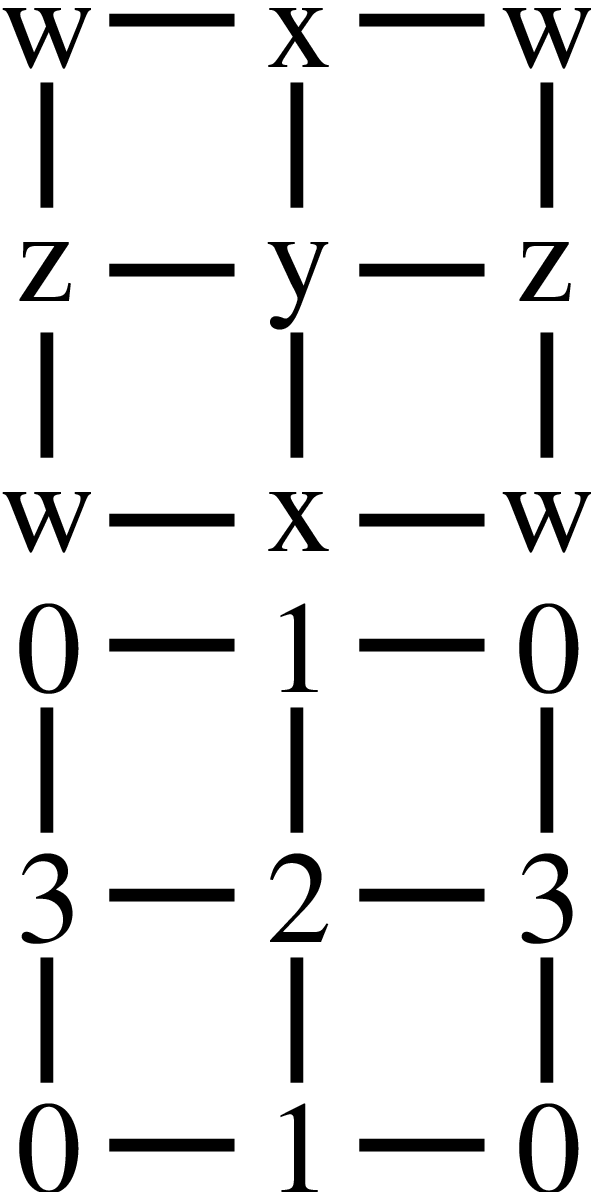}}
\hspace{0.2in}
\subfigure[``Electric" fields]{\includegraphics[width=0.21\textwidth]{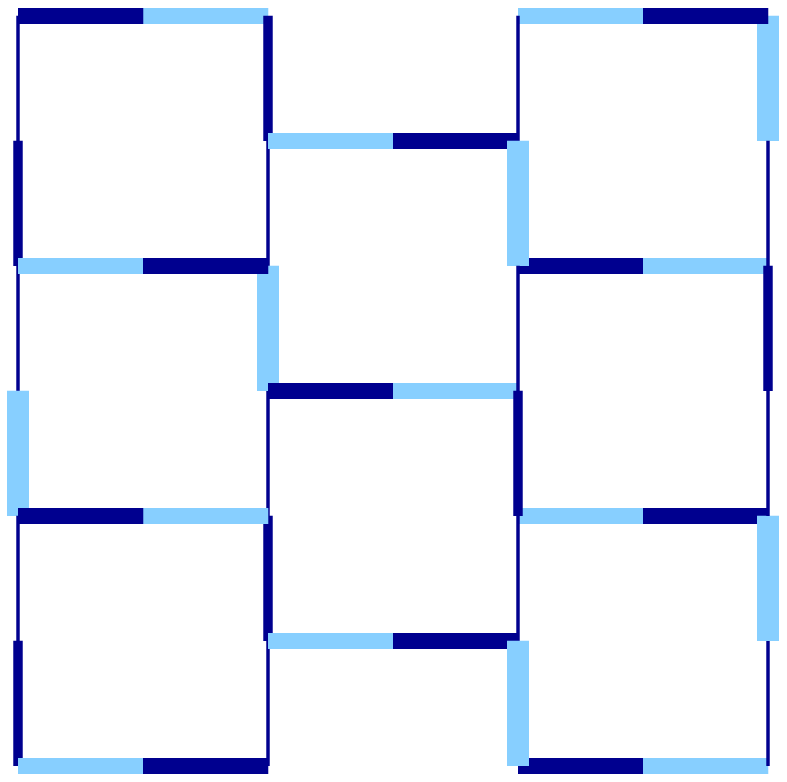}}
\caption{(a) The four sublattices $W$, $X$, $Y$, $Z$ of the square lattice. (b) the numerator of the square lattice fractional offsets $\zeta_v\in \{0,1/4,2/4,3/4\}$. (c) The electric fields $\langle e_p\rangle$. The thickness of the bonds (time-like plaquettes) represent the magnitude of $\langle e_p\rangle$ while the dark bonds have $\langle e_p\rangle>0$ and the light bonds have $\langle e_p\rangle<0$. The pattern of dark bonds has the same symmetry as the pin-wheel state.}
\label{fig:wxyz}
\end{figure}

To gain a simple understanding of the dual height model, consider softening the integer height fields, $h_v\to\chi_v$ such that $\chi_v$ is any real number. This necessarily introduces periodic potentials, $\cos2\pi\chi_v$, $\cos4\pi\chi_v$, etc. The resulting model is then $Z_\chi=\int {\mathcal D}\chi e^{-S}$ with
\begin{multline}\label{eq:sinegordon}
S =
 \frac{g}{2}\int_0^\beta d\tau \bigg[\sum_v a^2\big(\partial_\tau\chi_v\big)^2\!+\!
 \sum_{\langle vw\rangle} \big(\chi_v\!-\!\chi_w \big)^2  \\+\!
 2\sum_{\langle\langle vw\rangle\rangle} \big(\chi_v\!-\!\chi_w\big)^2\!+\!
 \sum_v M^2\cos\left(2\pi(\chi_v\!-\!\zeta_v)\right)\bigg]
\end{multline}
where for simplicity we have taken the continuum limit along the imaginary time axis and we have shifted $\chi_v$ by the fractional offsets $\zeta_v$. A simple saddle point analysis for this model has already been studied in Ref. \cite{Read:1990ys}. Since $\zeta_v$ takes on four different values depending on four different sublattices of the square lattice labeled by $W, X, Y, Z$ in Fig. \ref{fig:wxyz}, this saddle point analysis simply sets $\chi_v$ to one of $\{\chi_W,\chi_X,\chi_Y,\chi_Z\}$ depending on the sublattice and minimizes the energy with respect to these four parameters. Following Ref. \cite{Read:1990ys}, the result is 
\begin{eqnarray}\label{eq:heights}
  2\pi\chi_W &=& -\frac{\pi}{4} + \frac{M^2}{8\sqrt{2}} - \frac{M^4}{256}\\
  2\pi\chi_X &=& -\frac{\pi}{4} + \frac{M^2}{8\sqrt{2}} + \frac{M^4}{256}\\
  2\pi\chi_Y &=& -\frac{\pi}{4} -\frac{M^2}{8\sqrt{2}} - \frac{M^4}{256}\\
  2\pi\chi_Z &=& -\frac{\pi}{4} - \frac{M^2}{8\sqrt{2}} + \frac{M^4}{256}
\end{eqnarray}
so that the added terms on the diagonal bonds of the square lattice do not alter the ground state. 

Given these mean field parameters, we are left with the task of determining the physical properties of the ground state in terms of quantities defined on the direct lattice. To this end, we compute the static magnetic fields in the ground state $\langle b_{\ibar\jbar}\rangle \equiv \chi_{\ibar} - \chi_{\jbar}$ which are the gradient of the potential $\chi_{\ibar}$. As stated earlier, each of these dual magnetic fields correspond to an electric field $\langle b_{\ibar\jbar}\rangle =-\langle b_{\ibar\jbar}\rangle \to \langle e_p\rangle$ on a plaquette of the direct lattice. Spatial plaquettes have a vanishing $\langle e_{p}\rangle$, while temporal plaquettes have either $\langle e_p\rangle\!>\!0$, $\langle e_p\rangle\!=\!0$ or $\langle e_p\rangle\!<\!0$.  A plot of $\langle e_p\rangle$ is shown in Fig. \ref{fig:wxyz}b and has the same symmetry as the pin-wheel state shown in Fig. 2 of the main text.


\end{document}